\documentclass[letter,longauth,structabstract]{aa} % for the letters 
\usepackage{graphicx}
\usepackage{txfonts}
\begin{document}
\def\HII{H\,{\sc{ii}}}
\def\HI{H\,{\sc{i}}}
   \title{Star formation triggered by H\,{\sc{ii}} regions in our Galaxy}

   \subtitle{First results for N49 from the $\it{Herschel}$\thanks{Herschel is an ESA space observatory with science
instruments provided by European-led Principal Investigator
consortia and with important participation from NASA.} infrared survey of the Galactic plane}

   \author{A. Zavagno\inst{1}
\and L.D. Anderson \inst{1} 
\and D. Russeil \inst{1}
\and L. Morgan \inst{2}
\and G.S. Stringfellow \inst{3}
\and L. Deharveng\inst{1}
\and J.A. Rod\'on \inst{1}
\and T.P. Robitaille \inst{4}
\and J.C. Mottram \inst{5}
\and F. Schuller \inst{6}
\and L. Testi\inst{7}
\and N. Billot \inst{8} 
\and S. Molinari \inst{9} 
\and A. di Gorgio \inst{9}
\and J.M. Kirk\inst{10}
\and C. Brunt \inst{5}
\and D. Ward-Thompson\inst{10}
\and A. Traficante\inst{11}
\and M. Veneziani\inst{11} 
\and F. Faustini\inst{12} 
\and L. Calzoletti\inst{12}
          }

   \institute{Laboratoire d'Astrophysique de Marseille (UMR 6110 CNRS \& 
Universit\'e de Provence), 38 rue F. Joliot-Curie, 13388 Marseille 
Cedex 13, France
              \email{annie.zavagno@oamp.fr}
\and Astrophysics Research Institute, Liverpool John Moores
University, Twelve Quays House, Egerton Wharf, Birkenhead CH41 1LD
\and Center for Astrophysics and Space Astronomy Department of Astrophysical and Planetary Sciences 389 UCB Boulder Colorado
\and Spitzer Postdoctoral Fellow, Harvard-Smithsonian Center for
Astrophysics, 60 Garden Street, Cambridge, MA, 02138, USA
\and School of Physics, University of Exeter, Stocker Road, Exeter, EX4 4QL, UK
\and Max-Planck Institut f\"{u}r Radioastronomie, 
	Auf dem H\"{u}gel 69, D-53121 Bonn, Germany
\and ESO, Karl Schwarzschild-Strasse 2, D-85748 Garching 
	bei M\"{u}nchen, Germany
\and NASA Herschel Science Center, IPAC, Caltech, Pasadena, CA 91125, USA
\and INAF-IFSI, Fosso del Cavaliere 100, 00133 Roma, Italy
\and School of Physics and Astronomy, Cardiff University, Queens Buildings, The Parade, Cardiff, CF24 3AA, UK 
\and Dipartimento di Fisica, Università di Roma 2 "Tor Vergata", Rome, Italy
\and ASI Science Data Center, I-00044 Frascati (Rome), Italy
             %\email{annie.zavagno@oamp.fr}
             }

   \date{Received March 31, 2010; accepted April 12,2010}

% \abstract{}{}{}{}{} 
% 5 {} token are mandatory
 
  \abstract
   {It has been shown that by means of different physical mechanisms the expansion of H\,{\sc{ii}} regions can trigger the formation of new stars of all masses. This process may be important to the formation of massive stars but has never been quantified in the Galaxy. }
   {We use $\it{Herschel}$-PACS and -SPIRE images from the $\it{Herschel}$ Infrared survey of the Galactic plane, Hi-GAL, to perform this study.}
   {We combine the $\it{Spitzer}$-GLIMPSE and -MIPSGAL, radio-continuum and sub-millimeter surveys such as ATLASGAL with Hi-GAL to study Young Stellar Objects (YSOs) observed towards Galactic \HII\ regions. We select a representative \HII\ region, N49, located in the field centered on $l$=30$\degr$ observed as part of the Hi-GAL Science Demonstration Phase, to demonstrate the importance Hi-GAL will have to this field of research. }
   {Hi-GAL PACS and SPIRE images reveal a new population of embedded young stars, coincident with bright ATLASGAL condensations. The Hi-GAL images also allow us, for the first time, to constrain the physical properties of the newly formed stars by means of fits to their spectral energy distribution. Massive young stellar objects are observed at the borders of the N49 region and represent second generation massive stars whose formation has been triggered by the expansion of the ionized region.}
   {The first Hi-GAL images obtained using PACS and SPIRE have demonstrated the capability to investigate star formation triggered by \HII\ regions. With radio, submillimeter, and shorter wavelength infrared data from other surveys, the Hi-GAL images reveal young massive star-forming clumps surrounding the perimeter of the N49 \HII\ generated bubble. Hi-GAL enables us to detect a population of young stars at different evolutionary stages, cold condensations only being detected in the SPIRE wavelength range. The far IR coverage of Hi-GAL strongly constrains the physical properties of the YSOs. The large and unbiased spatial coverage of this survey offers us a unique opportunity to lead, for the first time, a global study of star formation triggered by \HII\ regions in our Galaxy. }


   \maketitle
%
%________________________________________________________________

\section{Introduction}
Ionized (H\,{\sc{ii}}) regions are known to trigger the formation of stars by means of various physical mechanisms (see Elmegreen~\cite{elm98} and Deharveng et al.~\cite{deh05} for a review). Several H\,{\sc{ii}} regions have been studied individually in the context of 
triggered star formation, focusing on the associated neutral material and the young stellar population (Zavagno et al.~\cite{zav06}; Deharveng et al.~\cite{deh09}; Pomar\`es et al.~\cite{pom09}; Bieging et al.~\cite{bie09}). These studies have shown that the expansion of H\,{\sc{ii}} regions can trigger the formation of new stars of all masses. The $\it{Spitzer}$-GLIMPSE survey of the Galactic plane (Benjamin et al.~\cite{ben03}) detected nearly 600 bubbles (Churchwell et al. \cite{chu06}). Deharveng et al. (\cite{deh10}) selected a series of 102 ionized bubbles and studied the star formation in 
their surroundings using $\it{Spitzer}$-GLIMPSE and MIPSGAL (Carey et al. \cite{car09}), radio (MAGPIS; Helfand et al. \cite{hel06} and VGPS; Stil et al. \cite{sti06}), and ATLASGAL (Schuller et al. \cite{sch09}) data. 
They show that 86\% of these bubbles enclose H\,{\sc{ii}} regions, and that more than 20\% of 64 bubbles (for which the ATLASGAL angular resolution is sufficient to resolve the spatial distribution of cold dust) show massive star formation on their borders. This indicates that triggering is important in the creation of massive stars and that hot photodissociation regions (PDRs) are a good place to study the earliest phases of massive-star formation.  

Our long-term aim is to use Hi-GAL (Molinari et al.~\cite{mol10}), combined with existing infrared, submillimeter, and radio surveys, to study the influence of H\,{\sc{ii}} regions on triggering the formation of stars in our Galaxy. Hi-GAL's extended wavelength coverage towards the far-infrared and its unprecedented sensitivity offer a unique opportunity to detect an embedded population of young sources that are not detected at shorter wavelengths. This allows us to observe intermediate and high-mass YSOs over the complete range of evolutionary stages. The unprecedented resolution of Hi-GAL also offers the opportunity to accurately characterize the physical nature of the sources by means of a detailed fit to their spectral energy distribution (SED).

In this Letter we study the bubble-shaped ionized region, N49, from the Churchwell et al.~(\cite{chu06}) catalogue to illustrate the purpose of our project. The N49 bubble was studied by Watson et al.~(\cite{wat08}) and by Deharveng et al.~(\cite{deh10}). However, these studies had no information in the 70--500\,$\mu$m range. This information, obtained with the PACS and SPIRE data presented here, allows us to discuss the star formation of this region in detail.    
 
\section{Hi-GAL results on N49}
\begin{figure*}[ht]
   \centering
   \includegraphics[angle=0,width=180mm]{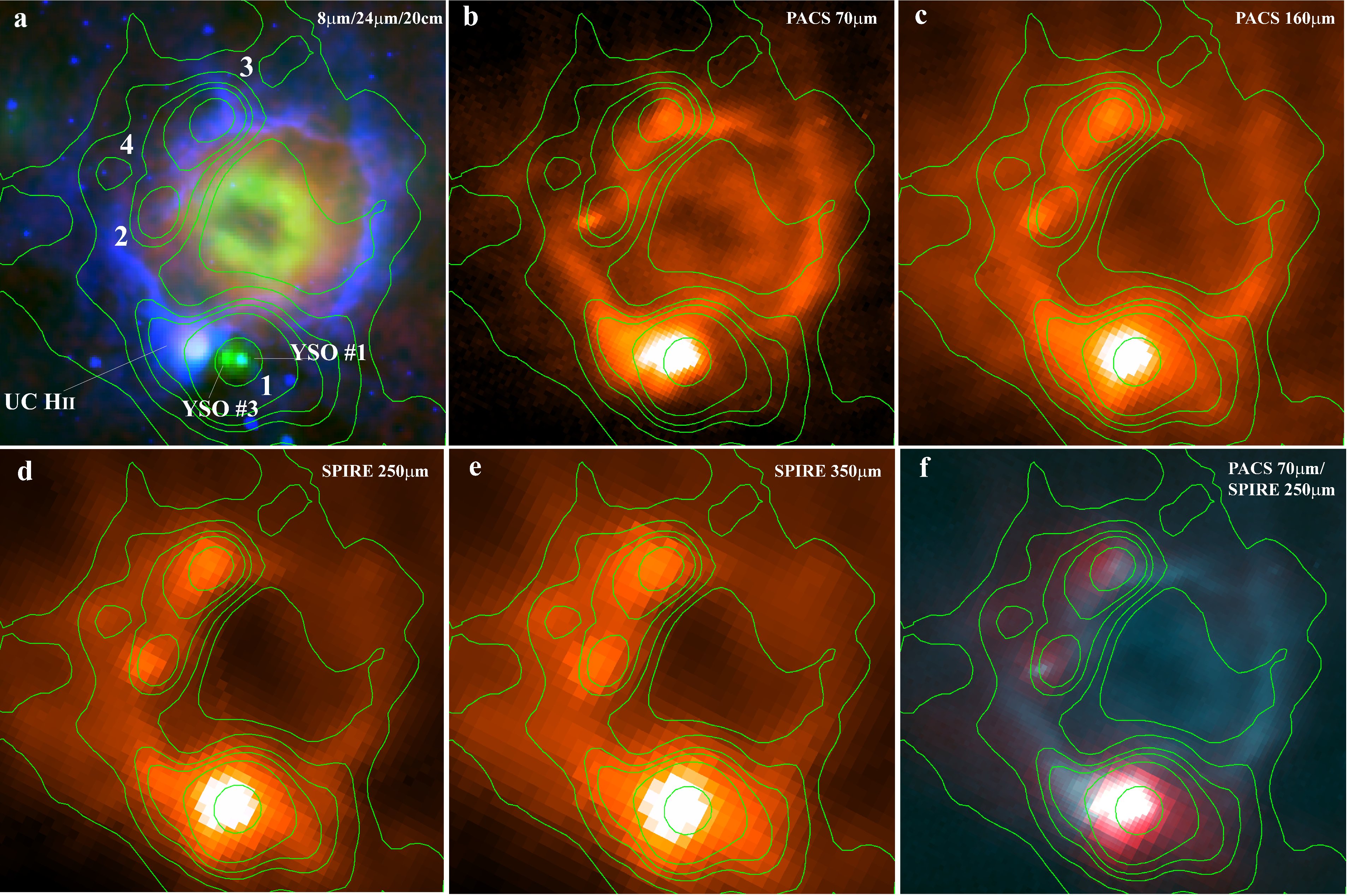}
   \caption{N49: a) Three-color composite image with 8\,$\mu$m {\it Spitzer}-GLIMPSE (blue), 24~$\mu$m MIPSGAL (green), 20~cm MAGPIS (red). ATLASGAL 870~$\mu$m contours are superimposed on all images (levels are 0.01, 0.1, 0.22, 0.31, 0.5, and 2~Jy/beam). ATLASGAL condensations 1 to 4 are identified. We have identified the two massive stage I YSOs (YSO\#1 and YSO\#3) discussed by Watson et al.~(\cite{wat08}). The ultracompact \HII\ region is also identified; b) PACS 70\,$\mu$m image; c) PACS 160\,$\mu$m image; d) SPIRE 250\,$\mu$m image; e) SPIRE 350\,$\mu$m image; f) Two-color composite image with PACS 70\,$\mu$m (blue) and SPIRE 350\,$\mu$m image (red). The field is 5$\arcmin\,\times$ 5$\arcmin$, centered on $l$=28.83$\degr$, $b$=$-$0.23$\degr$
  }
              \label{N49}%
    \end{figure*}

We use PACS and SPIRE images obtained in parallel mode at 70, 160, 250, 350 and 500\,$\mu$m (at resolutions of 5\arcsec, 11.4\arcsec, 17.9\arcsec, 25\arcsec and 35.7\arcsec, respectively) to determine the influence of \HII\ regions on their surroundings using 
the N49 \HII\ bubble as an example. Our general goal is to quantify the efficiency of \HII\ regions on triggering star formation around the perimeter of their radiation-driven expanding bubble.
%--------------------
Figure~\ref{N49} shows images of this region at different wavelengths, including the new PACS and SPIRE images. 
ATLASGAL 870\,$\mu$m contours are superimposed on all images. 
% Figure
This \HII\ region is centered on $l$=28.83$\degr$, $b$=$-$0.23$\degr$ and located at 5.5~kpc (Anderson et al. \cite{and09}). It has a diameter of 4~pc. A compact \HII\ region lies on its border (see Fig.~\ref{N49}a).  ATLASGAL 870\,$\mu$m emission shows a fragmented shell of collected material surrounding the ionized region (green contours on Fig.~\ref{N49}). Three condensations are seen surrounding the ionized region and a secondary peak is detected farther away. These are referred to as condensations 1 through 4, respectively, and are indicated in Fig.~\ref{N49}a.
Using ATLASGAL data, Deharveng et al.~(\cite{deh10}) derived a mass for the entire shell of 4250~M$_{\odot}$, and the mass of the brightest condensation in this shell (condensation 1 in Fig.~\ref{N49}a) of 2350~M$_{\odot}$. The NH$_3$ velocity measurements show that the condensations 1, 2, and 3 are associated with the ionized region (Wienen et al., $\it{in\,preparation}$). No measurement exists for condensation 4 but $^{13}$CO (1-0) data from the Galactic Ring survey (Jackson et al.~\cite{jac06}) show that it is also associated with N49. Two young stellar objects (identified YSO\#1 and YSO\#3 in Fig.~\ref{N49}a) are seen as bright 24\,$\mu$m sources towards condensation 1, the brightest at 870\,$\mu$m. These two YSOs are massive and in an early evolutionary stage (Watson et al. \cite{wat08}). Using high resolution 6.7~GHz methanol maser measurements, Cyganowski et al.~(\cite{cyg09}) suggest that a rotating disk is associated with YSO\#3.
An ultracompact \HII\ region nearby YSO\#3 is shown in Fig.~\ref{N49}a as a bright, but more diffuse, source at 24\,$\mu$m. According to Deharveng et al.~(\cite{deh10}), this region may have been ionized by a B0V star. 

Figure~\ref{N49}b shows the PACS 70\,$\mu$m image with  superimposed ATLASGAL 870\,$\mu$m contours. We see that YSO\#3 dominates the emission at longer wavelengths. The ultracompact \HII\ region, bright at 24\,$\mu$m on the MIPSGAL image, is not bright in the 70-500\,$\mu$m range. A 70\,$\mu$m source is detected towards condensation 2 and 3. These sources are not detected at 24\,$\mu$m.  

Figure~\ref{N49}d shows the SPIRE 250\,$\mu$m image with ATLASGAL 870\,$\mu$m contours superimposed. The sources previously detected at 70\,$\mu$m  towards condensations 2 and 3 are seen. Weak emission is observed towards condensation 4. This emission was not detected at 70\,$\mu$m and 160\,$\mu$m. 

The PACS images clearly show the hot PDR that surrounds the \HII\ region. The 70\,$\mu$m and the 8\,$\mu$m emission of the PDR are very similar. The only difference comes from the PACS 70\,$\mu$m source seen towards condensation 2 and observed as a dark spot on the 8\,$\mu$m image. This indicates that condensation 2 and the PACS IR source are probably located in front of the PDR. The situation is different for condensation 3, for which no clear absorption is seen at 8\,$\mu$m suggesting that the condensation is located behind the PDR. The two color-composite image shown in Fig.~\ref{N49}f, PACS 70\,$\mu$m (blue) and SPIRE 350\,$\mu$m (red), indicate that parts of the region where star formation is occurring are denser or colder or both, as indicated by the red emission.       

The brightest IR source observed around N49, probably corresponding to YSO\#3, is located along an IR dark cloud seen on the 8\,$\mu$m GLIMPSE image. Since blobs are observed along this filament on larger scales, the origin of the formation of this bright IR source is questionable. Was it produced by the fragmentation and collapse of the shell surrounding N49 by means of the collect and collapse process (Elmegreen \& Lada~\cite{elm77}), or by the compression of a pre-existing blob by the N49 (and the ultracompact) \HII\ region(s)? Morphological considerations favor the first option as a pre-existing clump would have distorted the ionization front, which is not what we observe here. We discuss the star formation observed around N49 in the framework of the collect and collapse model in Sect.~3.      

\subsection{The spectral energy distribution fitting}
We used the on-line SED fitting tool of Robitaille et al.~(\cite{rob07}) to derive the physical properties of the sources observed towards ATLASGAL condensations bordering N49. Fluxes were measured at 70, 160, 250, 350, 500, and 870\,$\mu$m by rebinning all the images to the PACS 70\,$\mu$m resolution, using a square aperture of 100$\arcsec$, 50$\arcsec$, 70$\arcsec$, and 40$\arcsec$ for condensations 1, 2, 3, and 4, respectively (2.67, 1.33, 1.87, and 1~pc for the N49 distance of 5.5~kpc), and subtracting emission from a local background. 

Watson et al.~(\cite{wat08}) identified seven YSOs towards N49. Their YSO\#5 and \#6, observed near condensation 4, are probably not related because they are not observed at longer wavelengths on the $\it{Herschel}$ images. Watson et al.~(\cite{wat08}) did not detect the red sources observed with PACS and SPIRE towards the ATLASGAL condensations 2 and 3. As seen in their Fig.~17, their SED fits are not constrained at wavelengths beyond 24\,$\mu$m due to a lack of data. The Hi-GAL coverage between 70 and 500\,$\mu$m, together with the ATLASGAL 870\,$\mu$m emission, strongly constrain the SED of the observed sources, allowing us to derive their physical properties. 

In Table~1, we present our results of the SED fitting for all sources, taking into account fits with $\chi^2-\chi^2_{best}$ per datapoint $ < 3 $. 
\begin{table}[h]
\caption{Results of the SED fitting}
\begin{tabular}{l r c c c}
  \hline
               &      &      &    &    \\   
  & $M_{\rm star}$  & 
   $\dot{M}_{\rm envelope}$ & $M_{\rm envelope}$ & $L$ \\
     & ($M_{\sun}$)  & (10$^{-3}$ $M_{\sun}~yr^{-1}$) & ($M_{\sun}$)&  (10$^3$ $L_{\sun}$) \\
  \hline
Cond.~1 &  20--30 & 8--10 & 5000  & 50\\
Cond.~2 & 8--10  & 4 & 8 & 1  \\
Cond.~3 & 8--15 & 2--5 & 2000--5000 & 2--5  \\
\hline
\end{tabular}
\end{table}
%-------------------------

\begin{figure}
   \centering
   \includegraphics[angle=0,width=90mm]{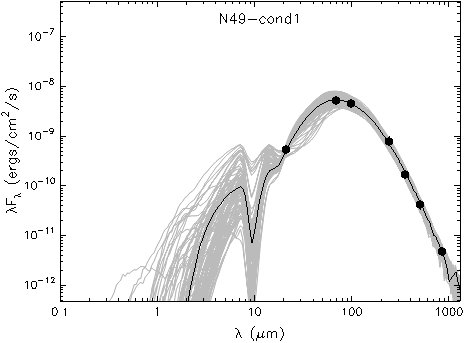}
   \caption{Results of the fit to the fluxes measured towards condensation 1 in N49 using the on-line SED fitting tool of Robitaille et al.~(2007). The best-fit (dark solid line) is strongly constrained by the Hi-GAL wide wavelength coverage. The grey lines show all the range of models. The 24\,$\mu$m flux is taken from Watson et al.~(2008).} 
              \label{fit}%
    \end{figure}

Towards condensation 1, a high luminosity, high mass source is found that has a high accretion rate. The values given in Table~1 are for the entire condensation 1, including YSO\#1 and YSO\#3. The 70\,$\mu$m emission is dominated by YSO\#3.  This is an extended green object and is associated with a methanol maser (Cyganowski et al.~\cite{cyg08}, \cite{cyg09}), which implies a massive YSO with an age younger than 3.5$\times$10$^{4}$ years (Breen et al.~\cite{bre10}). However, no emission is seen in its direction at 20~cm on the MAGPIS image. The high accretion rate derived for condensation~1 (see Table~1) may prevent the development of the ionized region (Churchwell~\cite{chu02}). The emission may also be optically thick at 20~cm. 
The results of the fit for condensation~1 are shown in Fig.~\ref{fit}. 
% Figure

The sources observed towards condensations 2 and 3 are also massive objects with high accretion rates.  
The emission that is barely detected at 250\,$\mu$m towards condensation 4 cannot be fit using the Robitaille et al. SED fitting tool because no data at shorter wavelengths exist to constrain the fit. This object is probably a cold core with no internal source of heating. However, it is interesting to detect this emission farther away from the ionization front as it may represent an earlier evolutionary stage in the star formation process. The 870\,$\mu$m ATLASGAL peak flux of 0.26~Jy/beam indicates a column density of N(H$_{2}$)=  6.8$\times$10$^{21}$~cm$^{-2}$ (2.2$\times$10$^{22}$~cm$^{-2}$) for a temperature of 20~K (10~K). This represents a low density condensation compared to the values derived by Deharveng et al.~(\cite{deh10}) around Galactic bubbles where star formation is clearly observed.        

In all cases, the age of the source is not well constrained by the fitting procedure. Other indicators, such as chemical clocks, outflows, and masers, are needed to more tighly constrain the ages.       
\section{Discussion}
Following the analytical approach of Whitworth et al.~(\cite{whi94}, see their Sect.~5), we discuss the observed star formation around N49 in the context of the collect and collapse process. The present radius of N49 (2~pc), its estimated age (0.5--1~Myr), and the spectral type of its ionizing star (O5V, Watson et al.~\cite{wat08}, Everett \& Churchwell~\cite{eve10}) allow us to derive a mean value for the initial density of the surrounding medium, n$_0 \simeq$~10$^4$~cm$^{-3}$. 
Assuming this density and following the formula given by Whitworth et al.~(\cite{whi94}), we can estimate the time at which the fragmentation begins, $t_{\rm{fragment}} \simeq$0.5~Myr and the radius and column density of the shell at this time, $R_{\rm{fragment}} \simeq$1.55~pc and $N_{\rm{fragment}} \simeq$1.6$\times$10$^{22}$~cm$^{-2}$, which are similar to the values we derived towards condensation 4. All these values strongly depend on the sound speed in the shocked gas a$_s$ (which is between 0.2 and 0.6 km/s) and n$_0$, the initial density of the medium. The values we give here are derived using a$_s$=0.2 km/s and n$_0$=10$^4$ cm$^{-3}$. If n$_0$ is lower, we obtain a $t_{\rm{fragment}}$ higher than the estimated age for N49, which is not possible because the fragments are already seen. The same happens if we increase a$_s$. However, the first set of values that provide acceptable results for the start time of the fragmentation and radius (and column density) of the shell at that time does not provide acceptable results for the mass of the fragments, which are too low (around 7.5~M$_{\odot}$) compared to the masses derived by Deharveng et al.~(\cite{deh10}) using the ATLASGAL measurements (of about 2500~M$_{\odot}$, 240~M$_{\odot}$, and 355~M$_{\odot}$ for condensations 1, 2 and 3, respectively). As discussed below, the wind dynamics that exists towards N49 can modify this simplified approach. A finer age determination for the sources observed towards the ATLASGAL condensations may help in more tightly constraining this model.    

The Hi-GAL images and results of the SED fits show that five massive YSOs (YSO\#1 and \#3, the UC \HII\ region, and IR sources towards condensations 2 and 3) are formed (or are in the process of forming) on the border of N49. This region is ionized by an O5V star and stellar winds are believed to play a crucial role in the formation of this bubble. Everett \& Churchwell~(\cite{eve10}) studied in detail the case of N49 as a wind-blown bubble. They conclude that the survival of dust within the ionized gas in the presence of winds requires a high density in the surrounding medium. This high density together with greater dynamical motion caused by stellar winds might have increased the efficiency of triggered star formation, in terms of number and mass of the second-generation stars. Other dusty wind-blown bubbles will be observed in Hi-GAL and the unbiased access to the YSOs population will help us to understand whether winds play a crucial role in increasing the efficiency of triggered star formation.   

\section{Conclusions}
We have presented the first PACS and SPIRE images from Hi-GAL of the bubble-shaped Galactic \HII\ region N49. This region is used as an illustration of the study dedicated to the star formation triggered by Galactic \HII\ regions that we plan to lead for the whole survey, combining the Hi-GAL results with other infrared and radio surveys of the Galactic plane. We have shown that:  
   \begin{itemize}
      \item{The Hi-GAL SPIRE and PACS images allow us to study the distribution of young sources towards N49. The far-IR fluxes have been measured and strongly constrain the spectral energy distribution of these sources. This allows us to characterize their properties. }      
     \item{The PACS images reveal the existence of red young stellar objects towards two ATLASGAL condensations, sources that had not been previously detected at shorter wavelengths. The ultracompact \HII\ region is not clearly seen in the $\it{Herschel}$ range. The bright YSO\#3 Watson et al. observed to be coincident with condensation 1 dominates the emission at longer wavelengths.}
      \item{SED fits for the 3 sources detected by $\it{Herschel}$ towards millimeter condensations using the Robitaille et al.~(\cite{rob07}) model show that these sources are young and massive. However, their age has not been constrained and other indicators are needed to refine the discussion of star formation history in this region.}
	\item{Five massive stars are forming in the N49 PDR. The high star formation efficiency in N49 may be due to the presence of winds from the first generation massive star.} 
   \end{itemize}
The study of N49 using a multiwavelength approach shows that Hi-GAL enables measurements for the crucial far-IR range to be made that are essential to infer the properties of the YSOs. Hi-GAL clearly provides important insight into star formation triggered by expanding \HII\ regions. Seventy-six \HII\ regions of all shapes are detected in the $l$=30$\degr$ field and a higher density of Hi-GAL sources is clearly observed towards these regions, indicating that star formation triggered by \HII\ regions may be an important process.    

\begin{acknowledgements}
    Part of this work was supported by the ANR (\emph{Agence Nationale pour la Recherche}) project
      "PROBeS", number ANR-08-BLAN-0241 (LA).
\end{acknowledgements}

\end{document}